\documentclass{Interspeech}

\title{A Two-Step Learning Framework for Enhancing Sound Event Localization and Detection}

\author[affiliation={1}]{Hogeon}{Yu}

\affiliation{}{Hyundai Motor Company}{Republic of Korea}
\email{hogeon.yu@hyundai.com}


\usepackage{comment}

\begin{document}
\interspeechcameraready
\maketitle

\begin{abstract}
Sound Event Localization and Detection (SELD) is crucial in spatial audio processing, enabling systems to detect sound events and estimate their 3D directions.
Existing SELD methods use single- or dual-branch architectures: single-branch models share SED and DoA representations, causing optimization conflicts, while dual-branch models separate tasks but limit information exchange.
To address this, we propose a two-step learning framework.
First, we introduce a tracwise reordering format to maintain temporal consistency, preventing event reassignments across tracks. 
Next, we train SED and DoA networks to prevent interference and ensure task-specific feature learning.
Finally, we effectively fuse DoA and SED features to enhance SELD performance with better spatial and event representation.
Experiments on the 2023 DCASE challenge Task 3 dataset validate our framework, showing its ability to overcome single- and dual-branch limitations and improve event classification and localization.
\end{abstract}

\section{Introduction}

The Sound Event Localization and Detection (SELD) system detects sound events in three-dimensional space, tracks their temporal activity, and estimates their direction of arrival. These systems are widely used in real-world applications, including crime detection, smart home monitoring, and human-robot interaction, enhancing situational awareness and rapid response.
However, real-world environments pose challenges due to overlapping sound events, making it difficult for machines to localize and classify sounds.
While humans naturally perceive and differentiate sounds, machines require advanced sound event detection (SED) and direction of arrival (DoA) techniques to achieve comparable performance.

The DCASE challenge, first introduced in 2019\cite{0}, has played a key role in advancing SELD research, initially focused on static single-source scenarios using synthetic multichannel audio generated from mono audio files with impulse responses. Over time, the challenge has evolved to encompass more complex acoustic environments, incorporating moving sources, diverse impulse responses, and real spatial sound scenes with polyphonic and overlapping events under low SNR conditions.

Various deep learning-based approaches have been proposed to tackle these challenges.
Existing SELD networks can be broadly categorized into single-branch and dual-branch architectures.
Early research primarily employed a single network with two separate output branches\cite{1,2,3}.
However, this often leads to optimization conflicts, as SED and DoA tasks learn different feature representations, causing mutual interference during training\cite{4,5}.
To overcome these challenges, dual-branch networks allow each task to learn independent features while sharing complementary information\cite{6,7,8}. 
Another approach integrates SED and DoA into a single-branch structure, learning both tasks simultaneously.\cite{9,10,11,12}.
However, these approaches still have certain limitations. While the dual-branch strategy leverages soft-parameter sharing to enhance optimization by sharing network weights between SED and DoA, it may dilute task-specific information due to parameter sharing. The single-branch networks may struggle to effectively capture the relationship between SED and DoA features.

Inspired by human auditory perception, our approach builds on previous methods while addressing their limitations.
The human auditory system has the ability to focus on sound occurring in a specific direction and recognize the type of event based on it\cite{23}. Additionally, even with a single auditory signal (e.g., hearing with one ear), humans can quickly determine the type of event and later, using both binaural signals (e.g., hearing with both ears), they refine the localization of the sound source and extract spatial information.
Building on these insights, our approach refines existing SELD models through a two-step learning strategy.
This approach builds on our previous work, which demonstrated the benefits of structured DoA and SED guidance\cite{24}. By separately training both tasks and then integrating them strategically, we enhance model robustness and generalization.


\begin{figure*}
    \centering
    \includegraphics[width=1\textwidth]{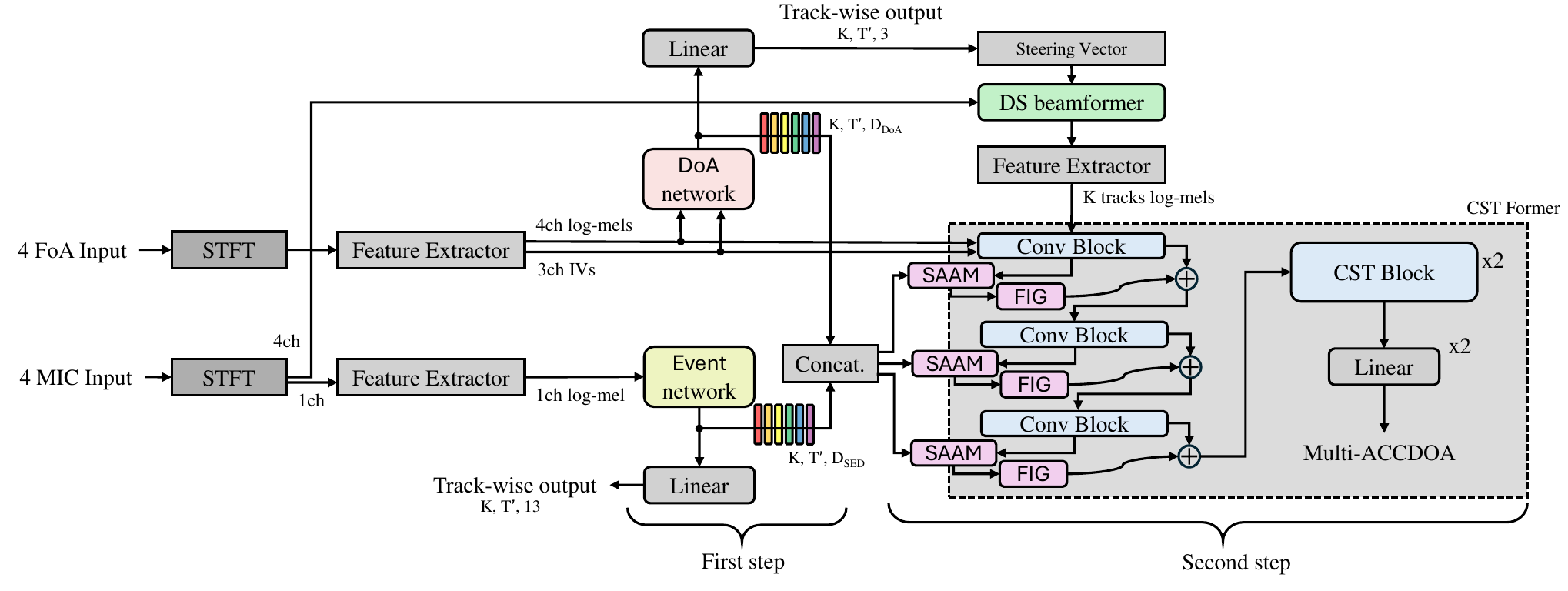}
    \caption{Overview of the proposed SELD network. The network separately processes DoA and SED features before fusion. $K$ was set to 6, and the feature dimensions of $D_{SED}$ and $D_{DoA}$ of SED and DoA were both set to 64. $T'$ corresponds to the number of frames of the target labels.}
    \label{fig:seld}
\end{figure*}

\section{Proposed Method}
Our proposed method follows a two-step learning framework to optimize SELD performance.
First, we independently train DoA and SED networks to prevent interference. We adopt IPDnet\cite{13} and ATST-SED\cite{14} for DoA and SED, extending their architectures to fit the polyphonic SELD scenario.
We propose a trackwise output format to prevent event type reassignment across tracks, ensuring temporal consistency. Unlike conventional formats\cite{4,5,6,7}, where events may shift between tracks over time, our method maintains stable track allocation.
These modifications not only increase the stability of the SED and DoA network but also enable the computation of steering vectors for individual event sources in each track\cite{26,27}, facilitating beamforming-based signal enhancement.
In the second step, we enhance SELD performance by fusing the pre-trained DoA and SED features. 
To achieve this, we adopt the multimodal fusion method from the LAVT framework\cite{15}, enabling DoA and SED features to complement each other effectively. The overall architecture of our method is illustrated in Figure~\ref{fig:seld}.

\subsection{Step 1: DoA and SED Estimation Network}

\subsubsection{Estimating the position of polyphonic sound sources}

We adopt IPDnet, which estimates the direct path Inter-channel Phase Difference (DP-IPD) to predict sound source locations. 
IPDnet which consists of full-band and narrow-band networks, efficiently captures inter-frequency correlations and temporal dependencies while tracking multiple moving sound sources simultaneously.

\begin{figure*}
    \centering
    \includegraphics[width=0.85\textwidth]{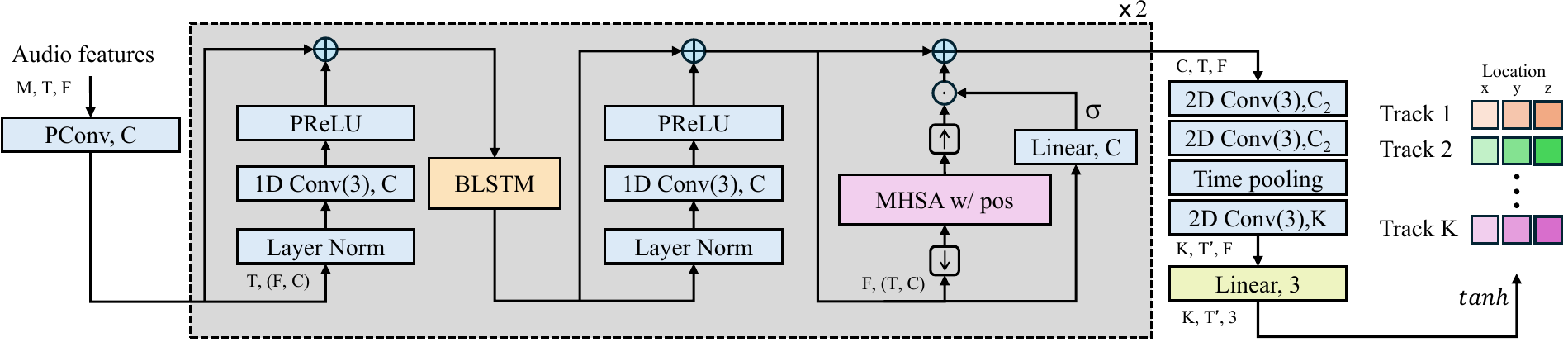}
    \caption{Architecture of the DoA network in the first step. The network processes audio features with input dimensions of $M \times T \times F$, where $M$ is the number of input channels, $T$ is the time frames, and $F$ is the frequency bins. The input dimensions for each block are indicated before it, formatted as "batch size × (dimensions of a single sample within the batch)". \(\downarrow\) and \(\uparrow\) in full-band denote downsampling with average pooling and upsampling with nearest interpolation. Each convolution operation is specified in the format: $(kernel \ size), feature \ maps$.}
    \label{fig:doa}
\end{figure*}

To enhance performance, we expand the network architecture with additional modifications. Unlike IPDnet, which applies residual connections after full-band and narrow band layers, we replace them with PConv at the input stage. In the full-band network, we apply frequency-convolution modules before and after the full-band BLSTM to strengthen correlations between adjacent frequencies. This approach improves localization stability and minimizes inter-frequency information loss.
Inspired by \cite{25}, we replace LSTM with Multi-Head Self-Attention (MHSA) with positional encoding (PE) in the narrow-band network, applying temporal down-sampling before MHSA and up-sampling afterward. While LSTM sequentially propagates information across frames, MHSA dynamically attends to relevant frames, making it more effective for capturing discontinuous sound events. The addition of PE ensures that temporal dependencies are preserved. To maintain consistency between the predicted DoA information and ground-truth labels, the down-sampled resolution is matched with the temporal resolution of the target, reducing computational complexity while preserving global acoustic features.
This proposed structure is illustrated in Figure~\ref{fig:doa}.

However, IPDnet requires accurate phase differences between microphones as a target for training, but the DCASE challenge Task3 dataset, \textit{synth-set}\cite{16}, does not provide distance information between the sound source and microphones.
To address this limitation, we substitute DP-IPD with the relative location $(x,y,z)$ enabling direct regression of spatial coordinates. We utilize log-mel spectrograms and intensity vectors (IVs) from FoA array signals to predict the cartesian coordinates of up to $K$ overlapping events per frame, denoted as \( Y_{k}^{DoA} \in \mathbb{R}^{T' \times 3} \), \( k \in \{1, ..., K\} \), where \(K\) represents the maximum number of tracks.

\subsubsection{Estimating events of the polyphonic sound sources}

In this work, we adopt ATST-SED, a high-performing network for SED tasks, to detect polyphonic sound events.
ATST-SED extracts precise frame-level audio features with high temporal resolution, making it well-suited for downstream tasks in other SED domains.
For SELD tasks, it is crucial to consider the possibility of detecting multiple instances of the same event simultaneously.
Although conventional SED networks\cite{17,18}, such as ATST-SED, support polyphonic event detection, they are not inherently structured to capture multiple occurrences of the same event at the same time.
To address this limitation, we extend the trackwise output format, enabling the detection of identical events across multiple tracks, and thereby improving the model’s ability to represent overlapping instances of the same event.
To implement this extension, we modify the final linear layer to produce an output of $n_{\mathrm{class}} \times n_{\mathrm{tracks}}$ per frame.
Additionally, since no extra unlabeled data is used in this experiment, we freeze the pre-trained feature extractor instead of fine-tuning it.
Finally, our SED network takes the log-mel spectrograms of the first MIC array signal as input for the network and estimates K overlapping locations of event per frame. The final output of the network is expressed as follows.
 \( Y_{k}^{SED} \in \mathbb{R}^{T' \times n_{\mathrm{class}}} \), \( k \in \{1, ..., K\} \), where $n_{\mathrm{class}}$ represents the number of event class.

\subsubsection{Reordering track-wise target label}

We introduce a novel trackwise reordering format during training, which ensures that each track maintains a consistent event type over a defined audio clip, preserving temporal consistency.
In conventional trackwise formats, sound events could be dynamically reassigned to different tracks over time. 
For example, if the maximum number of tracks is three, and “female,” “male,” and “music” events appear in 1 second, with “male” assigned to track 1, the “music” event could later be reassigned to track 1 if “male” ends before 2 seconds.
To ensure more stable event tracking, our proposed trackwise format prevents such dynamic track reassignment, ensuring that each track preserves the same event type throughout a given interval.
This approach enhances the alignment between DoA and SED by maintaining spatial orientation clarity and ensuring temporal continuity of events.
Additionally, by using the trackwise directional output from the DoA network, we compute steering vectors for beamforming.

\subsection{Step 2: DoA and Event Guidance Method for the SELD system}

\subsubsection{Delay-and-Sum Beamforming}
The phase difference between microphones is calculated using the sound source direction estimated for each track in the first step by the DoA network.
With the computed phase differences, we apply Delay and Sum (DS) beamforming, which enhances signals arriving from target directions while attenuating diffuse noise. 
The distance between estimated source coordinate \( Y_{k}^{DoA} \) and the microphone array coordinate \( (x_{mic}^{m}, y_{mic}^{m}, z_{mic}^{m}) \) is calculated using the Euclidean distance formula:

\begin{multline}
d_{k,m}(t) = 
  \sqrt{(x_k(t)-x_{\mathrm{mic}}^m)^2 
    + (y_k(t)-y_{\mathrm{mic}}^m)^2} \\[+0.3em]
  +(z_k(t)-z_{\mathrm{mic}}^m)^2
\end{multline}
where \(m\) is microphone channel, \( x_{k}(t), y_{k}(t)\), \( z_{k}(t) \) are sound event coordinates at track \(k\), time \( t \).
Next, the steering vector \( s_{k,m}(f,t) \) is computed as:

\begin{align}
s_{k,m}(f,t) = \exp \left( -j \frac{2\pi f}{c} d_{k,m}(t) \right)
\end{align}
where \( f \) represents the frequency in Hz and  \( c \) is sound speed in air. To account for cases where an event is absent in certain time frames, we introduce a track-wise weighting factor $w_{k}(t)$, defind as follows:
\begin{align}
w_k(t) = \begin{cases}
\sqrt{x_k^2 + y_k^2 + z_k^2}, & \text{if } d_{k}(t) \geq 0.5 \\  0.01 & \text{otherwise} \end{cases}
\end{align}
where \(d_{k}(t)\) denote the source distance.
The final beamformed output for track $k$ is given by:

\begin{align}
Y_k(f,t) = \frac{1}{M} \sum_{m=1}^{M} w_k(t) s_{k,m}^{*}(f,t)X_m(f,t)
\end{align}
where \( X_m(f,t) \) is STFT coefficients of microphone and \(^{*} \) is complex conjugate and \(Y_k(f,t)\) represents the beamformed output for track k.
The beamformed outputs are then incorporated into the CNN block to refine spatial feature learning, enhancing both event detection and localization.

\subsubsection{Fusion method}

We propose a fusion method to effectively integrate the DoA and SED features extracted in the first step, constructing a unified representation for SELD network.
To achieve this, we apply the multi-modal fusion approach from LAVT, which initially incorporates the Pixel-Word Attention Module (PWAM) and Language Gate (LG).
In our SELD framework, these are adapted as the  Spatial-Acoustic Attention Module (SAAM) and Feature Interaction Gate (FIG) to facilitate spatial-acoustic feature integration.
Our method progressively integrates spatial-acoustic features across multiple CNN layers, allowing DoA and SED features to interact dynamically throughout the feature extraction process.
By incorporating SAAM and FIG within multiple CNN blocks, our approach enhances both feature interaction and localization accuracy.
We set the output channels of 1D convolution layers in SAAM to match the output dimensions of the CNN block, maintaining structural coherence.
\section{Experiments}

\subsection{Implementation Details}
For training, we utilized the STARSS23\cite{19} dev-set-train and a synthetic dataset, while evaluation was conducted on the dev-set-test.
The dataset provides two recording formats: first-order ambisonics (FOA) and microphone array (MIC).
The proposed model was trained using different input configurations for DoA and SED network. The DoA model utilized four log-mel spectrograms and three intensity vectors extracted from the FoA array as input, while the SED model used one log-mel spectrogram from the first channel of the MIC array. To preprocess the audio signals, we applied the STFT with a hop length of 0.02s and a window size of 0.04s to audio signals sampled at 24 kHz. The resulting spectrograms were then converted into log-mel representations using 64 mel filter banks to effectively capture spectral features for the DoA network. For the SED network, we extracted 128-dimensional log-mel features following the input configuration of the pretrained ATST-SED.
In the first step, the DoA and SED networks were trained with an initial learning rate of 1.0e-4 for a maximum of 250 epochs. The learning rate was decayed by a factor of 0.5 if the validation loss did not improve over three consecutive epochs. A batch size of 128 was used for training. We applied permutation-invariant training\cite{20} with mean squared error loss for DoA and binary cross-entropy loss for SED.
In the second step, we adopted the same preprocessing as CST-former\cite{12} to the four FoA log-mel spectrograms and three IVs,. The SELD network was trained for 500 epochs with 32 batch size. We use AdamW with an initial learning rate of 1.5e-3, fixed for the first 100 updates and followed by a linear decay. A 50-epoch warm-up scheduler was applied to stabilize training. The input signals were segmented into 5-second clips without overlap.

\subsection{Experiments Results}
\subsubsection{DoA results}

\begin{table}[th]
  \caption{DoA Network Performance Comparison}
  \label{tab:doa_result}
  \centering
  \begin{tabular}{ l  r  r  r  r }
    \toprule
    \multicolumn{1}{c}{\textbf{Model}} & 
    \multicolumn{1}{c}{\textbf{ACC}$\uparrow$} & 
    \multicolumn{1}{c}{\textbf{MDR}$\downarrow$} & 
    \multicolumn{1}{c}{\textbf{MAE}$\downarrow$} &  \\
    \midrule
    CST-Former  & $63.8$  & $34.2$  & $15.5$    \\
    IPDnet  & $67.1$    & $30.9$  & $16.3$    \\
    + spectral conv  & $67.8$  & $30.1$  & $16.8$  \\
    + temporal MHSA  & $70.1$   & $27.3$  & $17.6$   \\
    \bottomrule
  \end{tabular}
\end{table}

To evaluate the performance of our proposed DoA network enhancements, 
experiments were conducted using the same metrics as IPDnet.
In 2023 DCASE challenge Task 3, a detected event is considered correct if its localization error is within 20°.
Following this evaluation protocol, we set the threshold to 20° in our experiments.
Table 1 illustrates the performance improvements achieved by incorporating frequency convolution and replacing the narrow-band LSTM with temporal MHSA.
Experimental results demonstrate that integrating spectral convolution into the full-band network significantly enhances performance, resulting in higher accuracy (ACC) and a lower missed detection rate (MDR) compared to IPDnet.
Furthermore, replacing the narrow-band LSTM with temporal MHSA further improved ACC to 70.1\% and reduced the MDR to 27.3\%, demonstrating the effectiveness of leveraging both spectral and temporal dependencies.
While these modifications notably improved detection performance, they also led to a slight increase in mean absolute error (MAE), increasing slightly from 16.3° to 17.6°.
This trade-off suggests that the model achieves better event detection, though with a minor increase in angular estimation error. However, the error remains within the 20° threshold, confirming that the improved detection accuracy outweighs the minor compromise in localization precision.

\subsubsection{SED results}

\begin{table}[th]
  \caption{SED Network Performance Comparison}
  \label{tab:sed_result}
  \centering
  \begin{tabular}{ l c c}
    \toprule
    \textbf{Model} &
    \textbf{trackwise format} &
    \textbf{segemnt F$_{macro}$}\\
    \midrule
    CST-Former  & $\times$ & $0.386$      \\
    ATST-SED  & $\times$ & $0.542$    \\
    ATST-SED  & $\circ$ & $0.584$     \\
    \bottomrule
  \end{tabular}
\end{table}

The segment length was fixed at 100 ms, matching the temporal resolution in 2023 DCASE challenge Task 3.
We evaluated CST-former and ATST-SED and conducted experiments with two output formats: with and without the trackwise format. To ensure a fair comparison, we converted the trackwise ATST-SED output \((n_{\mathrm{class}}, n_{\mathrm{tracks}})\) to a non-trackwise format \(n_{\mathrm{class}}\) by taking the maximum values across tracks.
Table 2 shows that ATST-SED without trackwise format outperformed CST-Former, achieving an F-macro score of 0.542 compared to 0.386. This result indicates that ATST-SED is more effective in detecting sound events than CST-Former.
Applying the trackwise format further improved performance to 0.584.
This enhancement suggests that maintaining a consistent assignment between event classes and tracks helps reduce ambiguity and improve event continuity. These findings further confirm the benefits of trackwise modeling in polyphonic sound event detection.

\subsubsection{comparison with other method}

\begin{table}[th]
  \caption{Performance Comparison of SELD networks}
  \label{tab:model_performance}
  \resizebox{\linewidth}{!}{%
  \centering
  \setlength{\tabcolsep}{4pt} 
  \begin{tabular}{ l  c c c  c  c }
    \toprule
    \textbf{Model} & \textbf{ER$_{\leq 20^\circ}\downarrow$} & \textbf{F$_{\leq 20^\circ}\uparrow$} & \textbf{LE$_{CD}\downarrow$} & \textbf{LR$_{CD}\uparrow$} & \textbf{SELD$_{score}\downarrow$} \\
    \midrule
    2023 Baseline\cite{22}  & $0.57$  & $29.9$  & $22.0$  & $47.7$  & $0.4791$  \\
    DST attention\cite{21}  & $0.58$  & $39.5$  & $20.0$  & $55.8$  & $0.4345$  \\
    CST-former\cite{12}  & $0.56$  & $42.7$  & $17.9$  & $62.0$  & $0.4019$  \\
    MFF-EINV2\cite{7}  & $0.54$  & $42.5$  & $18.7$  & $62.6$  & $0.3980$  \\
    Ours \\ 
    + DoA fusion & $0.55$ & $42.8$ & $18.3$ & $62.1$ & 
    $0.4007$ \\  
    + BF & $0.54$  & $42.9$  & $17.9$  & $62.4$  & $0.3966$  \\
    + SED fusion & $0.54$  & $44.0$  & $18.4$  & $64.5$  & $0.3891$  \\
    \bottomrule
  \end{tabular}
  }
\end{table}
To evaluate the effectiveness of our proposed two-step approach, we conducted SELD performance experiments on the 2023 DCASE challenge task 3 dataset.
Table 3 presents a comparison of our model against state-of-the-art SELD networks, including CST-Former and MFF-EINV2.
Our approach integrates features in three configurations: + DoA fusion, which fuses only directional features; +BF, which additionally incorporates beamformed output; + SED fusion, which integrates all, including sound event features.
Our SELD network outperforms the 2023 baseline across all evaluation metrics, achieving an 18.8\% reduction in SELD score. While the DST attention network effectively captures spectro-temporal dependencies, our approach further improves the SELD score by 10.5\%, demonstrating its advantage in both localization and event classification.
In contrast, CST-Former and MFF-EINV2 exhibit strong performance in single-branch and dual-branch architectures, respectively, with reductions of 16.1\% and 16.9\% in SELD score from the baseline. 
However, our proposed method further refines this performance, achieving a 3.2\% improvement over CST-Former and a 2.2\% improvement over MFF-EINV2.
Furthermore, as shown in Table 3, our approach progressively enhances performance by integrating DoA and SED features, achieving the lowest SELD score of 0.3891. This confims the effectivess of our fusion method in improving SELD performance.


\section{Conclusion}
We propose a two-step learning framework for SELD, integrating structured fusion to leverage both SED and DoA features effectively. The trackwise reordering format and beamformed spatial features enhance temporal consistency and localization accuracy. Our model achieves the lowest SELD score on the 2023 DCASE Task 3 dataset, showing its effectiveness in SELD tasks.
In future work, we plan to extend DoA estimation for real-time operation and explore its application in developing more efficient SELD systems.



\bibliographystyle{IEEEtran}
\bibliography{mybib}

\end{document}